\documentclass[a4paper]{jpconf}
\usepackage{amsmath}
\usepackage{amsfonts, amsmath, amssymb,latexsym}
\usepackage{graphicx} 
\usepackage{xcolor,microtype}
\usepackage{subfig}
\usepackage{bm}
\usepackage{eucal}      

\usepackage{mathrsfs}
\usepackage{soul}
\usepackage{amsbsy}
\usepackage{psfrag}
\usepackage{mathtools}

\def\beq{\begin{equation}}
\def\eeq{\end{equation}}

\def\beq{\begin{equation}}
\def\eeq{\end{equation}}

\definecolor{olivegreen}{rgb}{0,0.6,0}

\def\spacce#1{\hskip #1pt}
\def\drawline#1#2{\raise 2.5pt\vbox{\hrule width #1pt height #2pt}}
\def\solid{\drawline{24}{.5}\nobreak}

\def\bdash{\hbox{\drawline{5.8}{.5}\spacce{2}}}

\def\dashed{\bdash\bdash\bdash\nobreak}

\def\circle{$\circ$\nobreak }

\def\trian{\raise 1.25pt\hbox{$\scriptstyle\triangle$}\nobreak}

\def\dtrian{\raise 1.25pt\hbox%
{$\scriptscriptstyle\bigtriangledown$}\nobreak}

\def\squar{\raise 1.25pt\hbox{$\scriptstyle\Box$}\nobreak}

\def\diamon{\raise 1.25pt\hbox{$\scriptstyle\diamond$}\nobreak}

\def\beq{\begin{equation}}
\def\eeq{\end{equation}}

\def\citalajim03{Del \'Alamo \& Jim\'enez (2003)}

\newcommand\newblock{\hskip .11em\@plus.33em\@minus.07em}

\begin{document}

\title{
  Alternative physics to understand wall turbulence: \\
  Navier--Stokes equations with modified linear dynamics}

\author{Adri\'an Lozano-Dur\'an}
\address{Center for Turbulence Research, Stanford University, CA 94305, USA}
\ead{adrianld@stanford.edu}

\author{Marios-Andreas Nikolaidis}
\address{Department of Physics, National and Kapodistrian University of Athens,  Athens 157 72, Greece}

\author{Navid C.~Constantinou}
\address{Research School of Earth Sciences and ARC Centre of Excellence for Climate Extremes, Australian National University,
  Canberra ACT 2601, Australia}

\author{Michael Karp}
\address{Center for Turbulence Research, Stanford University, CA 94305, USA}

\begin{abstract} 
   Despite the nonlinear nature of wall turbulence, there is evidence
   that the energy-injection mechanisms sustaining wall turbulence can
   be ascribed to linear processes. The different scenarios stem from
   linear stability theory and comprise exponential instabilities from
   mean-flow inflection points, transient growth from non-normal
   operators, and parametric instabilities from temporal mean-flow
   variations, among others. These mechanisms, each potentially
   capable of leading to the observed turbulence structure, are rooted
   in simplified theories and conceptual arguments. Whether the flow
   follows any or a combination of them remains unclear. In the
   present study, we devise a collection of numerical experiments in
   which the Navier-Stokes equations are sensibly modified to quantify
   the role of the different linear mechanisms. This is achieved by
   direct numerical simulation of turbulent channel flows with
   constrained energy extraction from the streamwise-averaged
   mean-flow.  We demonstrate that (i) transient growth alone is not
   sufficient to sustain wall turbulence and (ii) the flow remains
   turbulent when the exponential instabilities are suppressed. On the
   other hand, we show that (iii) transient growth combined with the
   parametric instability of the time-varying mean-flow is able to
   sustain turbulence.
\end{abstract}

\newpage

\section{Introduction: linear theories of self-sustaining wall turbulence}

Turbulence is a primary example of a highly nonlinear phenomenon.
Nevertheless, there is ample agreement that the energy-injection
mechanisms sustaining wall turbulence can be partially attributed to
linear processes~\cite{Jimenez2013}.  The different mechanisms have
their origins in linear stability theory \cite{Reynolds1972,
  Hamilton1995, Waleffe1997, Schoppa2002, Delalamo2006a,Hwang2010b,
  Hwang2011} and constitute the foundations of many control and
modeling strategies~\cite{Kim2006,Schmid2012}. Despite the ubiquity of
linear theories, the significance of different instabilities in fully
developed turbulence remains outstanding, and its relevance is
consequential to comprehend, model, and control the structure of
wall-bounded turbulence by linear methods (e.g.,
Refs.~\cite{Hogberg2003, Delalamo2006a, Hwang2010b, Morra2019}). Here,
we devise a collection of numerical experiments of turbulent flows
over a flat wall in which the Navier--Stokes equations are minimally
altered to suppress the energy transfer from the mean flow to the
fluctuating velocities via different linear instabilities.

\begin{figure}
 \begin{center}
   \includegraphics[width=0.95\textwidth]{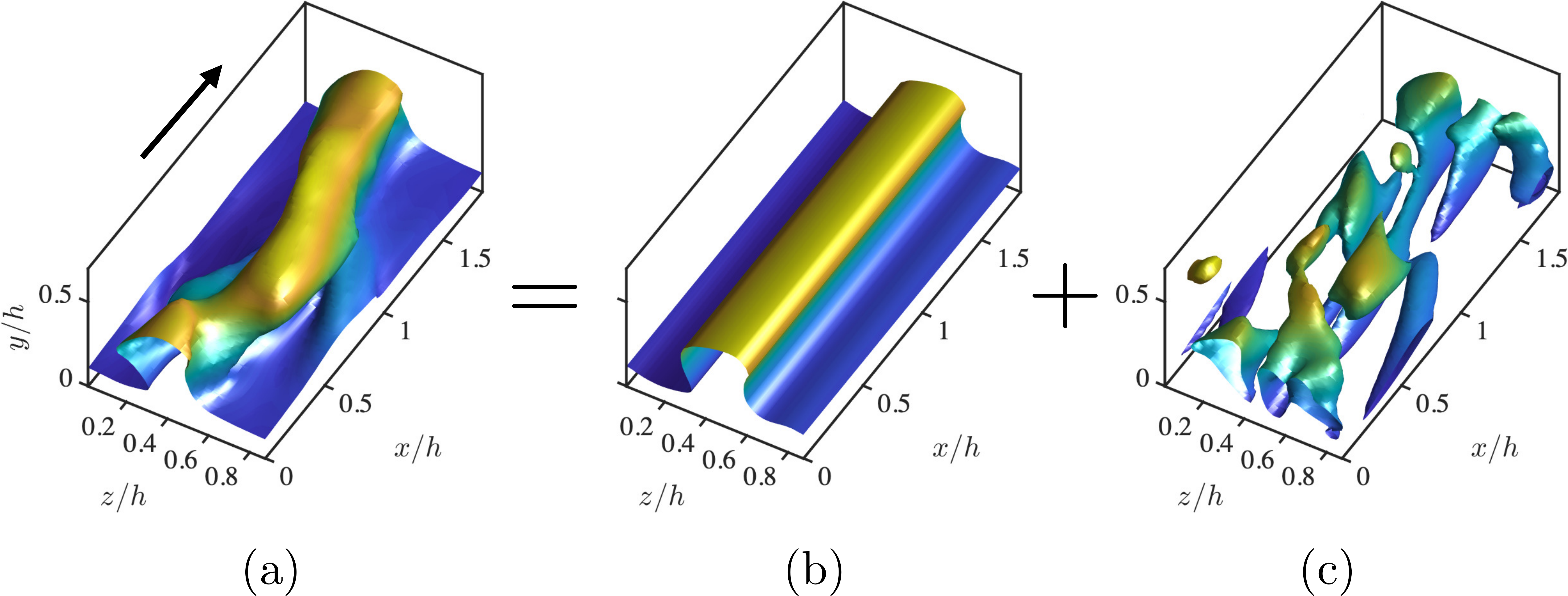}
 \end{center}
\caption{ Decomposition of the instantaneous flow into a streamwise
  mean base flow and fluctuations.  Instantaneous isosurface of
  streamwise velocity for (a)~the total flow $u$, (b)~the streak base
  flow $U$, and (c)~the absolute value of the fluctuations $|u'|$.
  The values of the isosurfaces are 0.8 (a and b) and 0.1 (c) of the
  maximum streamwise velocity.  Colors represent the distance to the
  wall located at $y=0$. The arrow in panel~(a) indicates the mean
  flow direction.
  \label{fig:snaphots}}
\end{figure}

Several linear mechanisms have been proposed within the fluid
mechanics community as plausible scenarios to rationalize the transfer
of energy from the large-scale mean flow to the fluctuating
velocities. Generally, it is agreed that the ubiquitous streamwise
rolls (regions of rotating fluid) and streaks (regions of low and high
streamwise velocity with respect to the mean)~\cite{Klebanoff1962,
  Kline1967} are involved in a quasi-periodic regeneration
cycle~\cite{Panton2001, Adrian2007, Smits2011, Jimenez2012,
  Jimenez2018,Lozano2019b} and that their space-time structure plays a
crucial role in sustaining shear-driven turbulence (e.g.,
Refs.~\cite{Kim1971, Jimenez1991, Butler1993, Hamilton1995,
  Waleffe1997, Schoppa2002, Farrell2012, Jimenez2012,
  Constantinou2014, Farrell2016, Lozano_brief_2018b}). Accordingly,
the flow is often decomposed into two components: a base flow defined
by some averaging procedure over the instantaneous flow, and the
three-dimensional fluctuations (or perturbations) about that base
flow.  In this manner, the ultimate cause maintaining turbulence is
conceptualized as the energy transfer from the base flow to the
fluctuating flow. Various base flows have been proposed in the
literature depending on the number of flow directions and period of
time to average the instantaneous flow. Here, we select as base flow
the instantaneous streamwise-averaged velocity $U(y,z,t)$ with zero
wall-normal ($V=0$) and spanwise ($W=0$) flow, where $y$ and $z$ are
the wall-normal and spanwise directions. Figure~\ref{fig:snaphots}
illustrates this flow decomposition.
%

The linear mechanisms proposed to explain the energy transfer from the
mean to the fluctuating flow can be categorized into: (i)~modal
inflectional instability of the mean streamwise flow, (ii)~non-modal
transient growth, and (iii)~non-modal transient growth assisted by
parametric instability of the time-varying mean streamwise flow.

In mechanism (i), it is hypothesized that the energy is transferred
from the mean profile $U(y,z,t)$ to the fluctuating flow
through a modal inflectional instability in the form of strong
spanwise flow variations~\cite{Hamilton1995, Waleffe1997}, corrugated
vortex sheets \cite{Kawahara2003}, or intense localized patches of
low-momentum fluid~\cite{Andersson2001, Hack2018}. Inasmuch as the
instantaneous realizations of the streaky flow are strongly
inflectional, the flow $U(y,z,t)$ is invariably unstable at a frozen
time $t$~\cite{Lozano_brief_2018b}. These inflectional instabilities
are markedly robust and their excitation has been proposed to be the
mechanism that replenishes the perturbation energy of the turbulent
flow~\cite{Hamilton1995, Waleffe1997, Andersson2001, Kawahara2003,
  Hack2014, Hack2018}. Consequently, the exponential instability of
the streak is thought to be central to the maintenance of wall
turbulence.

Mechanism (ii), transient growth, involves the redistribution of fluid
near the wall by streamwise vortices leading to the formation of
streaks via the Orr/lift-up mechanism \cite{Landahl1975, Farrell1993,
  Butler1993, Kim2000, Jimenez2012}.  In this case, the mean flow,
while exponentially stable, supports the growth of perturbations for a
period of time due to the non-normality of the linear operator that
governs the evolution of fluctuations. This process is referred to as
non-modal transient growth (e.g., Refs.~\cite{Butler1993,
  Delalamo2006a, Schmid2007, Cossu2009}). Additional studies suggest
that the generation of streaks is due to the structure-forming
properties of the linearized Navier--Stokes operator, independent of
any organized vortices \cite{Chernyshenko2005}, but the non-modal
transient growth is still invoked. The transient growth scenario
gained even more popularity since the work by Schoppa \& Hussain
\cite{Schoppa2002} (see also Ref. \cite{Giovanetti2017}), who argued
that transient growth may be the most relevant mechanism not only for
streak formation but also for their eventual breakdown. Schoppa \&
Hussain \cite{Schoppa2002} showed that most streaks detected in actual
wall-turbulence simulations are indeed exponentially stable. Instead,
the loss of stability of the streaks is better explained by transient
growth of perturbations that leads to vorticity sheet formation and
nonlinear saturation.

Finally, mechanism (iii) has been proposed in recent years by Farrell,
Ioannou and coworkers~\cite{Farrell2012, Farrell2016}. They adopted
the perspective of statistical state dynamics (SSD) to develop a
tractable theory for the maintenance of wall turbulence. Within the
SSD framework, the perturbations are maintained by an essentially
time-dependent, parametric, non-normal interaction of the fluctuations
and the streak, rather than by the inflectional instability of the
streaky flow discussed above (see also Ref.~\cite{Farrell2017}).

The scenarios (i), (ii), and (iii), although consistent with the
observed turbulence structure~\cite{Jimenez2018}, are rooted in
simplified theoretical arguments.  Whether the flow follows these or
any other combination of mechanisms for maintaining the turbulent
fluctuations is still to be established. In this study, we evaluate
the contribution of each linear mechanism by direct numerical
simulation of channel flows with constrained energy extraction from
the mean flow.

The study is organized as follows: Section~\ref{sec:numerical}
contains the numerical details of the simulations.  The results are
presented in Section~\ref{sec:results}, which is further subdivided
into three subsections each devoted to the investigation of one linear
mechanism. Finally, conclusions and future directions are offered in
Section~\ref{sec:conclusions}.

\section{Numerical experiments of minimal channel units}\label{sec:numerical}

To investigate the role of different linear mechanisms, we examine
data from spatially and temporally resolved simulations of an
incompressible turbulent channel flow driven by a constant mean
pressure gradient. Hereafter, the streamwise, wall-normal, and
spanwise directions of the channel are denoted by $x$, $y$, and $z$,
respectively, and the corresponding flow velocity components and
pressure by $u$, $v$, $w$, and $p$. The density of the fluid is $\rho$
and the channel height is $h$. The wall is located at $y=0$, where
no-slip boundary conditions apply, whereas free stress and no
penetration conditions are imposed at $y=h$. The streamwise and
spanwise directions are periodic. The grid resolution of the
simulations in $x$, $y$, and $z$ is $64 \times 90 \times 64$,
respectively, which is fine enough to resolve all the scales of the
fluid motion.

The simulations are characterized by the non-dimensional Reynolds
number, defined as the ratio between the largest and the smallest
length-scales of the flow, $h$ and $\delta_v = \nu/u_\tau$,
respectively, where $\nu$ is the kinematic viscosity of the fluid and
$u_\tau$ is the characteristic velocity based on the friction at the
wall~\cite{Pope2000}.  The Reynolds number selected is
$\mathrm{Re}_\tau = \delta/\delta_v \approx 180$, which provides a
sustained turbulent flow at an affordable computational
cost~\cite{Kim1987}.  In all cases, the flow is simulated for at least
$100 h/u_\tau$ units of time, which is orders of magnitude longer than
the typical lifetime of individual energy-containing
eddies~\cite{Lozano2014b}. The streamwise, wall-normal, and spanwise
sizes of the computational domain are $L_x^+ \approx 337$, $L_y^+
\approx 186$, and $L_z^+ \approx 168$, respectively, where the
superscript $+$ denotes quantities normalized by~$\nu$
and~$u_\tau$. Jimenez \& Moin \cite{Jimenez1991} showed that
turbulence in such domains contains an elementary flow unit comprised
of a single streamwise streak and a pair of staggered quasi-streamwise
vortices, that reproduce the dynamics of the flow in larger
domains. Hence, the current numerical experiment provides a
fundamental testbed for studying the self-sustaining cycle of wall
turbulence.

The simulations are performed with a staggered, second-order, finite
differences scheme \cite{Orlandi2000} and a fractional-step method
\cite{Kim1985} with a third-order Runge-Kutta time-advancing scheme
\cite{Wray1990}.  The solution is advanced in time using a constant
time step such that the Courant--Friedrichs--Lewy condition is below
0.5. The code has been presented in previous studies on turbulent
channel flows \cite{Lozano2016_Brief, Bae2018b, Bae2019}. The
streamwise and spanwise resolutions are $\Delta x^+\approx 6.5$ and
$\Delta z^+\approx3.3$, respectively, and the minimum and maximum
wall-normal resolutions are $\Delta y_{\mathrm{min}}^+\approx0.2$ and
$\Delta y_{\mathrm{max}}^+\approx6.1$.  All the simulations were run
for at least $100h/u_\tau$ after transients.

We focus on the dynamics of the fluctuating velocities
$\boldsymbol{u}' \equiv (u', v', w')$, defined with respect to the
streak base flow
\begin{equation}
U(y,z,t) \equiv \frac{1}{L_x} \int_{0}^{L_x} u(x,y,z,t) \,\mathrm{d}x,
\end{equation}
such that $u' \equiv u- U$, $v' \equiv v$, and $w' \equiv w$. We have
not included in the base flow the contributions from the streamwise
average $w$ and $w$ components, as is traditionally done in the study
of stability of the streaky flow~\cite{Reddy1993, Waleffe1997,
  Schoppa2002}. The fluctuating velocity vector $\boldsymbol{u}'
\equiv (u',v',w')$ is governed by
\begin{gather} \label{eq:NS}
\frac{\partial\boldsymbol{u}'}{\partial t} =
\mathcal{L}(U)\boldsymbol{u}'+ \boldsymbol{N}(\boldsymbol{u}'),
\end{gather}
where $\mathcal{L}$ is the linearized Navier--Stokes operator for the
fluctuating state vector about the instantaneous $U(y,z,t)$ (see
Figure~\ref{fig:snaphots}b) and $\boldsymbol{N}$ collectively denotes
the nonlinear terms (which are quadratic with respect to fluctuating
flow fields). Both $\mathcal{L}$ and $\boldsymbol{N}$ account for the
kinematic divergence-free condition $\boldsymbol{\nabla} \cdot
\boldsymbol{u}' = 0$. The corresponding equation of motion for
$U(y,z,t)$ is obtained by averaging the Navier--Stokes equations in
the streamwise direction.

We consider three numerical experiments. First, we simulate the
Navier--Stokes equations without any modification, in which the linear
instabilities of perturbations are naturally allowed. We refer to this
case as the ``regular channel''. The two additional experiments entail
a modification of the Navier--Stokes equations and they are described
in the sections below.

\section{Wall turbulence with constrained linear mechanisms}\label{sec:results}

\subsection{Wall turbulence without exponential instability of the streaks}
\label{subsec:nonmodal}

\begin{figure}
 \begin{center}
   \includegraphics[width=0.95\textwidth]{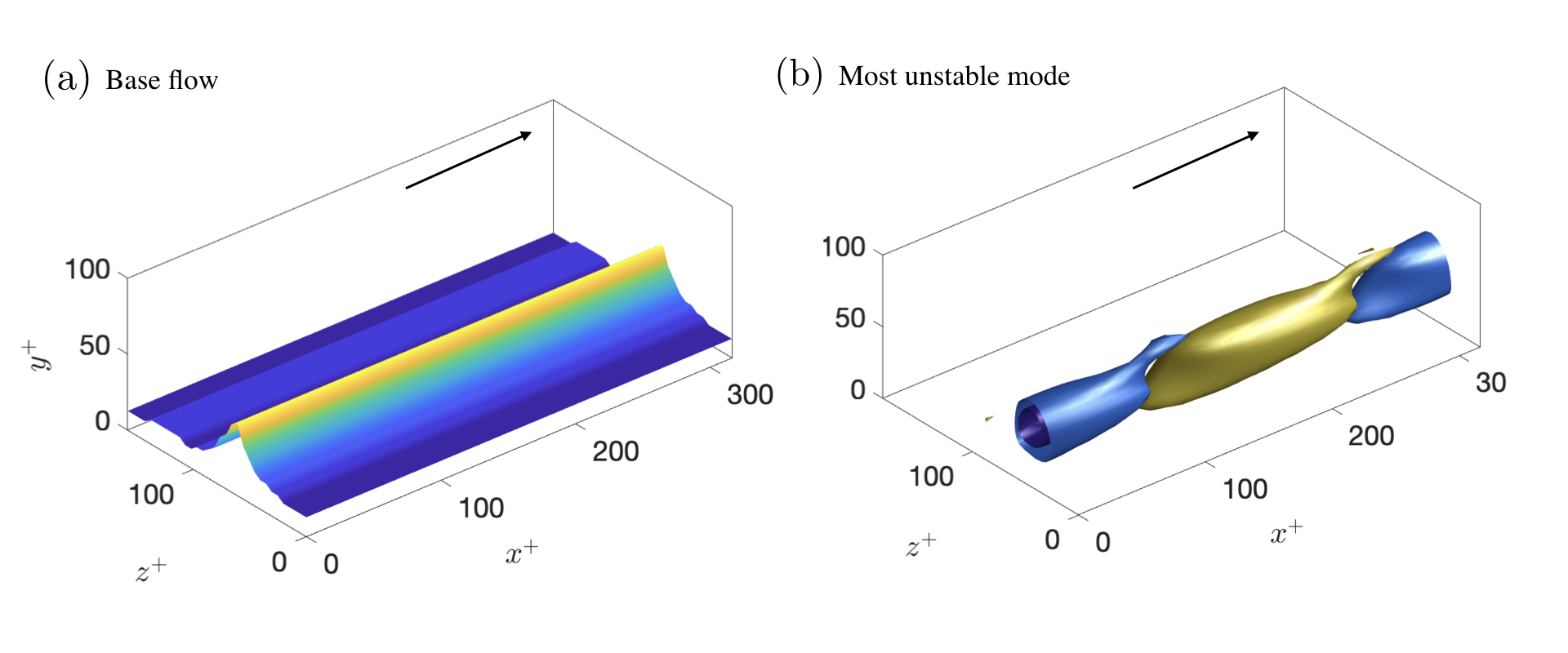}
 \end{center}
\caption{Representative exponential instability of the streak. (a)
  Instantaneous isosurface of the base flow $U$. The value of the
  isosurface is $0.8$ of the maximum. (b) Isosurface of the
  instantaneous streamwise velocity for the eigenmode associated with
  the most unstable eigenvalue $\lambda_\mathrm{max} h/u_\tau \approx
  3$ at $t=5.1 h/u_\tau$. The values of the isosurface are $-0.5$
  (blue) and $0.5$ (yellow) of the maximum streamwise
  velocity. \label{fig:modal_example} }
\end{figure}

The exponential instabilities of the instantaneous streamwise mean
flow at a given time are obtained by eigenanalysis of the matrix
representation of the operator $\mathcal{L}$ about the instantaneous
base flow $U$,
\begin{equation}
\mathcal{L}(U) = \mathcal{Q} \Lambda \mathcal{Q}^{-1},
\end{equation}
where $\mathcal{Q}$ consists of the eigenvectors organized in columns
and $\Lambda$ is the diagonal matrix of associated eigenvalues,
$\lambda_j + i\omega_j$. The base flow is unstable when any of the
growth rates $\lambda_j$ is positive.  Figure~\ref{fig:modal_example}
shows a representative example of the streamwise velocity of an
unstable eigenmode. The predominant eigenmode has the typical sinuous
structure of positive and negative patches of velocity flanking the
velocity streak side by side, which may lead to its subsequent
meandering and breakdown.
%

\begin{figure}
 \begin{center}
   \includegraphics[width=0.8\textwidth]{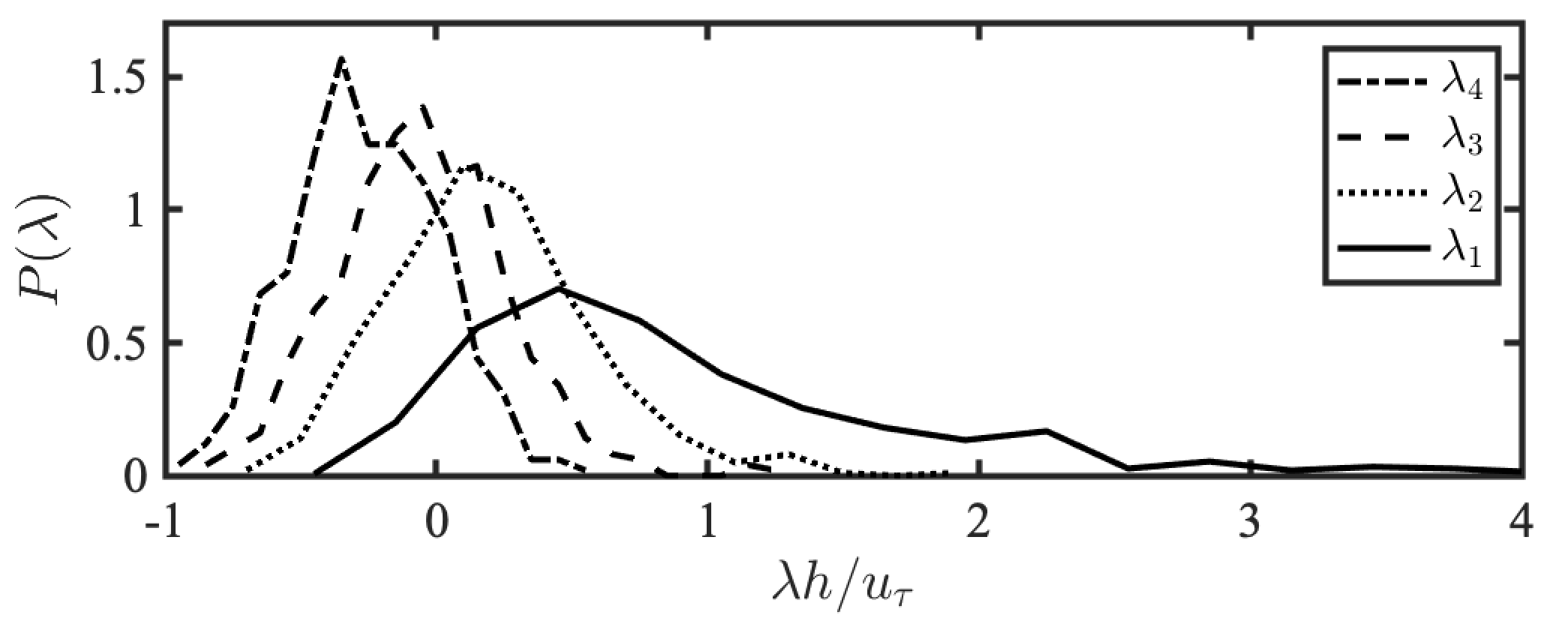}
 \end{center}
 \caption{ Probability density functions of the growth rate of the
   four least stable eigenvalues of $\mathcal{L}(U)$,
   $\lambda_1>\lambda_2>\lambda_3>\lambda_4$. \label{fig:P_eig}}
\end{figure}

\begin{figure}
 \begin{center}
 \includegraphics[width=0.95\textwidth]{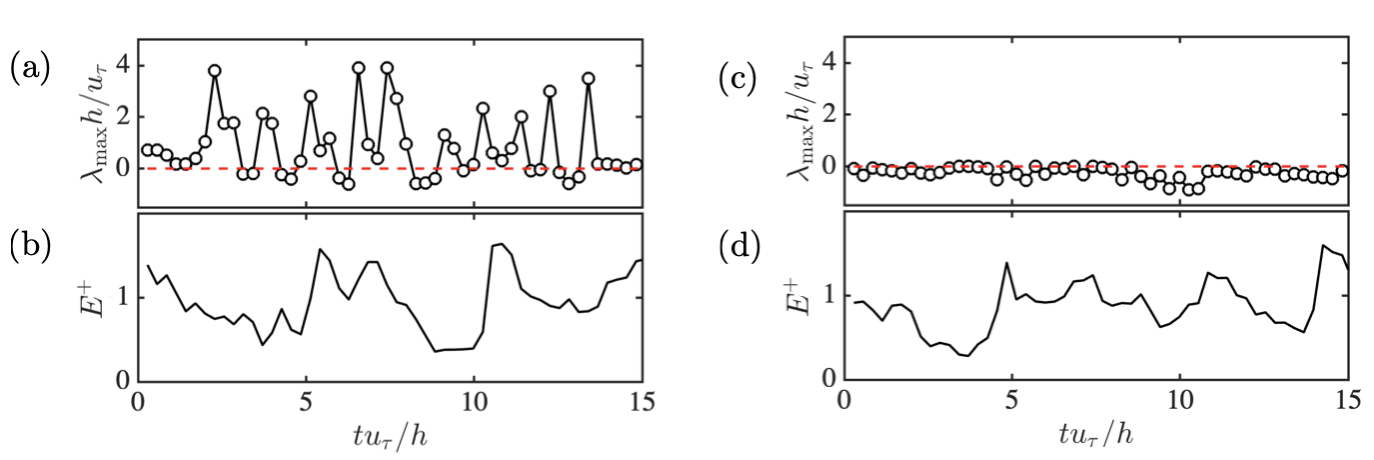}
 \end{center}
\caption{ (a,c) The time-history of the most unstable
  eigenvalue~$\lambda_{\rm max}$ of (a)~$\mathcal{L}$ for the regular
  channel flow and (c)~$\tilde{\mathcal{L}}$ for the channel flow with
  suppressed exponential instabilities. (b,d) The time-history of the
  kinetic energy of the perturbations~$E=\boldsymbol{u}' \cdot
  \boldsymbol{u}'/2$ averaged over the channel domain for (b)~the
  regular channel flow and (d)~for the channel flow with suppressed
  exponential instabilities.
 \label{fig:lambda}}
\end{figure}

Figure~\ref{fig:P_eig} shows the probability density functions of the
growth rate of the four least stable eigenvalues of $\mathcal{L}(U)$.
On average, the operator $\mathcal{L}$ contains 2 to 3 unstable
eigenmodes at any given instant. The time-history of the maximum
growth rate supported by $\mathcal{L}$, denoted by
$\lambda_{\mathrm{max}}$, is shown in Figure~\ref{fig:lambda}(a). The
flow is exponentially unstable ($\lambda_{\mathrm{max}}>0$) 70\% of
the time. The corresponding kinetic energy of the perturbations
averaged over the channel is shown in Figure~\ref{fig:lambda}(b).
%
%

Rigorously, we expect the linear instability to manifest in the flow
only when $\lambda_{\mathrm{max}}$ is much larger than time rate of
change of $U$, defined as $\lambda_U \equiv
(\mathrm{d}E_U/\mathrm{d}t)/(2E_U)$ with $E_U \equiv 1/(L_yL_z)
\int_{0}^{L_y} \int_{0}^{L_z} (1/2) U^2\,\mathrm{d}y \mathrm{d}z$ the
streak energy.  The ratio $\lambda_{\mathrm{max}}/\lambda_U$ for
$\lambda_{\mathrm{max}}>0$ is on average around $10$, i.e., the
time-changes of the streak $U$ are ten times slower than the maximum
growth rate predicted by the linear stability analysis.  Hence, the
exponential growth of disturbances is supported for a non-negligible
fraction of the flow history, and exponential instabilities stand as a
potential mechanism sustaining wall turbulence. Note that the argument
above does not imply that exponential instabilities are necessarily
relevant for the flow, but only that they could be realizable in terms
of characteristic time-scales.

For the second numerical experiment, we modify the operator
$\mathcal{L}$ so that all the unstable eigenmodes are rendered neutral
for all times. We refer to this case as the ``channel with suppressed
exponential instabilities'' and we inquire whether turbulence is
sustained in this case. The approach is implemented by replacing
$\mathcal{L}$ at each time-instance by the exponentially-stable
operator
\begin{equation}\label{Atilde}
\tilde{\mathcal{L}} = \mathcal{Q} \tilde{\Lambda} \mathcal{Q}^{-1},
\end{equation}
where $\tilde{\Lambda}$ is the stabilized version of $\Lambda$
obtained by setting the real part of all unstable eigenvalues of
$\Lambda$ equal to zero.  We do not modify the equation of motion for
$U(y,z,t)$.  The stable counterpart of $\mathcal{L}$ in
Eq.~\eqref{Atilde}, $\tilde{\mathcal{L}}$, represents the smallest
intrusion into the system to achieve exponentially stable wall
turbulence at all times while leaving other linear mechanisms almost
intact (we discuss this further in Section \ref{subsec:TG}).
Figure~\ref{fig:lambda}(c) shows the maximum modal growth rate of
$\tilde{\mathcal{L}}$ at selected times with the instabilities
successfully neutralized. It was verified that turbulence persists
when $\mathcal{L}$ is replaced by $\tilde{\mathcal{L}}$
(Figure~\ref{fig:lambda}d).

\begin{figure}
 \begin{center}
   \includegraphics[width=0.95\textwidth]{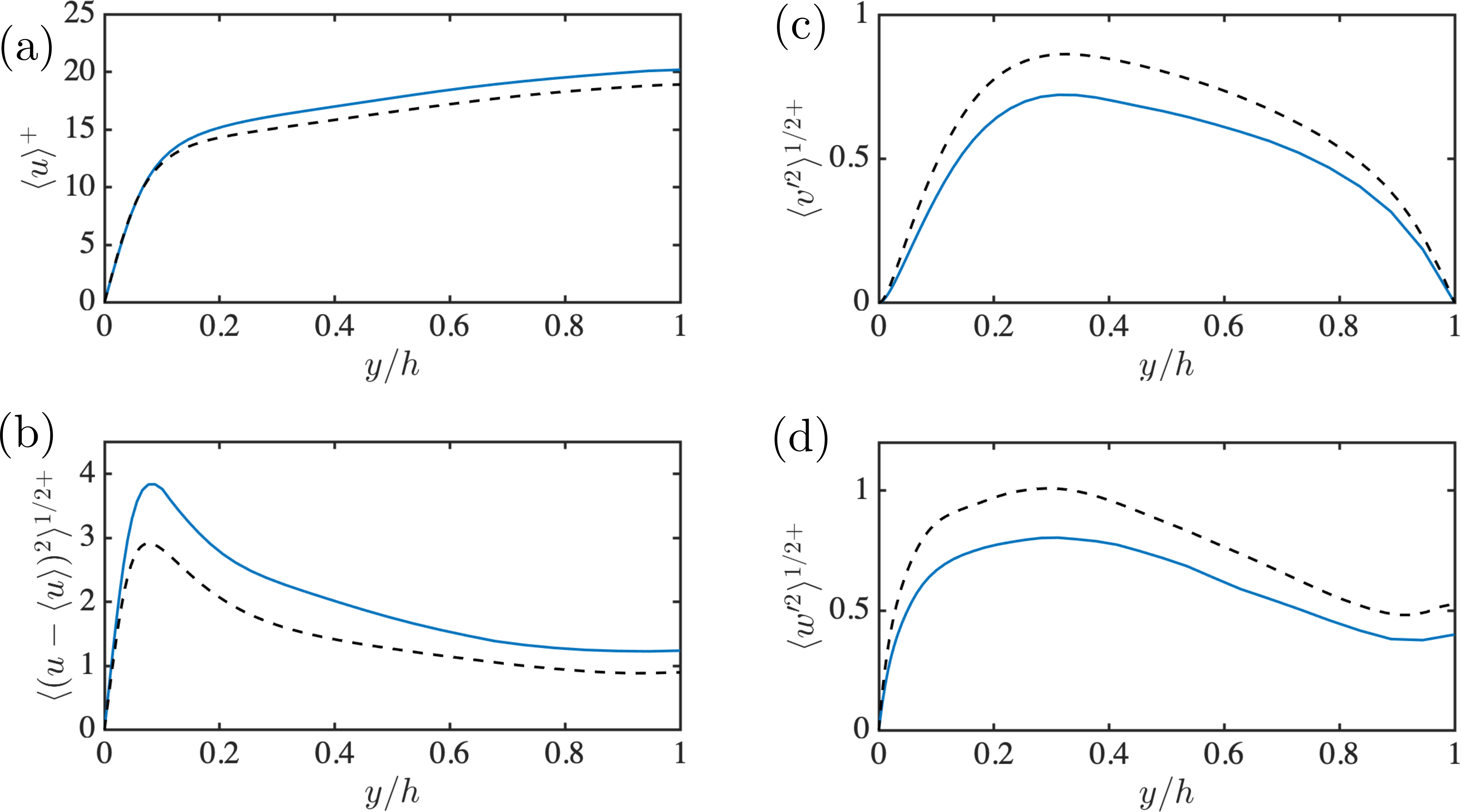}
 \end{center}
\caption{ (a) Streamwise mean velocity profile as a function of the
  wall-normal distance and (b) streamwise, (c) wall-normal, and (d)
  spanwise root-mean-squared fluctuating velocities for the regular
  channel (\dashed) and the channel with suppressed exponential
  instabilities (\textcolor{cyan}{\solid}). The Reynolds number of
  both simulations is $\mathrm{Re}_\tau = 186$. Angle brackets
  represent averaging in the homogeneous directions and time.
\label{fig:stats}}
\end{figure}

The main result of this section is presented in
Figure~\ref{fig:stats}, which compares the mean velocity profiles and
turbulence intensities for the regular channel and the channel with
suppressed exponential instabilities. The statistics are compiled for
the statistical steady state after any initial transients.  Notably,
the turbulent channel flow without exponential instabilities is
capable of sustaining turbulence.  The difference of roughly
15\%--25\% in the turbulence intensities between the cases indicates
that, even if the linear instability of the streak manifests in the
flow, it is not a requisite for maintaining turbulent
fluctuations. The new flow equilibrates at a state with augmented
streamwise fluctuations (Figure~\ref{fig:stats}b) and depleted cross
flow (Figure~\ref{fig:stats}c,d). The outcome is consistent with the
occasional inhibition of the streak meandering or breakdown via
exponential instability, which enhances the streamwise velocity
fluctuations, whereas wall-normal and spanwise turbulence intensities
are diminished due to a lack of vortices succeeding the collapse of
the streak.
%

\subsection{Wall turbulence exclusively supported by transient growth}
\label{subsec:TG}

We quantify the transient growth supported by $U(y,z,t_0)$ at a fixed
time $t_0$ as the mechanism energizing the fluctuating velocities. The
potential effectiveness of transient growth is characterized by the
optimal gain, $G$, defined as
\begin{equation}
G(t=T) = \frac{\boldsymbol{u}'^\dagger \exp(\mathcal{L}^\dagger T) \exp(\mathcal{L}T) \boldsymbol{u}' }{\boldsymbol{u}'^\dagger \boldsymbol{u}'},
\end{equation}
where $(\cdot)^\dagger$ denotes conjugate transpose, and $T$ is the
time-horizon for optimal gain. The square of the largest singular
value of the linear propagator
\begin{equation}
\exp(\mathcal{L}T) = \mathcal{M} \Sigma \mathcal{N}^\dagger,
\end{equation}
provides the maximum gain, $G_\mathrm{max}$, where the columns of
$\mathcal{M}$ and of $\mathcal{N}$ are the input modes (or
left-singular vectors) and output modes (or right-singular vectors) of
$\exp(\mathcal{L}T)$, respectively, and $\Sigma$ is a diagonal matrix,
whose entries are the singular values of $\exp(\mathcal{L}T)$ denoted
by $\sigma_j$ \cite{Butler1993,Farrell1996}.

\begin{figure}
 \begin{center}
   \includegraphics[width=0.95\textwidth]{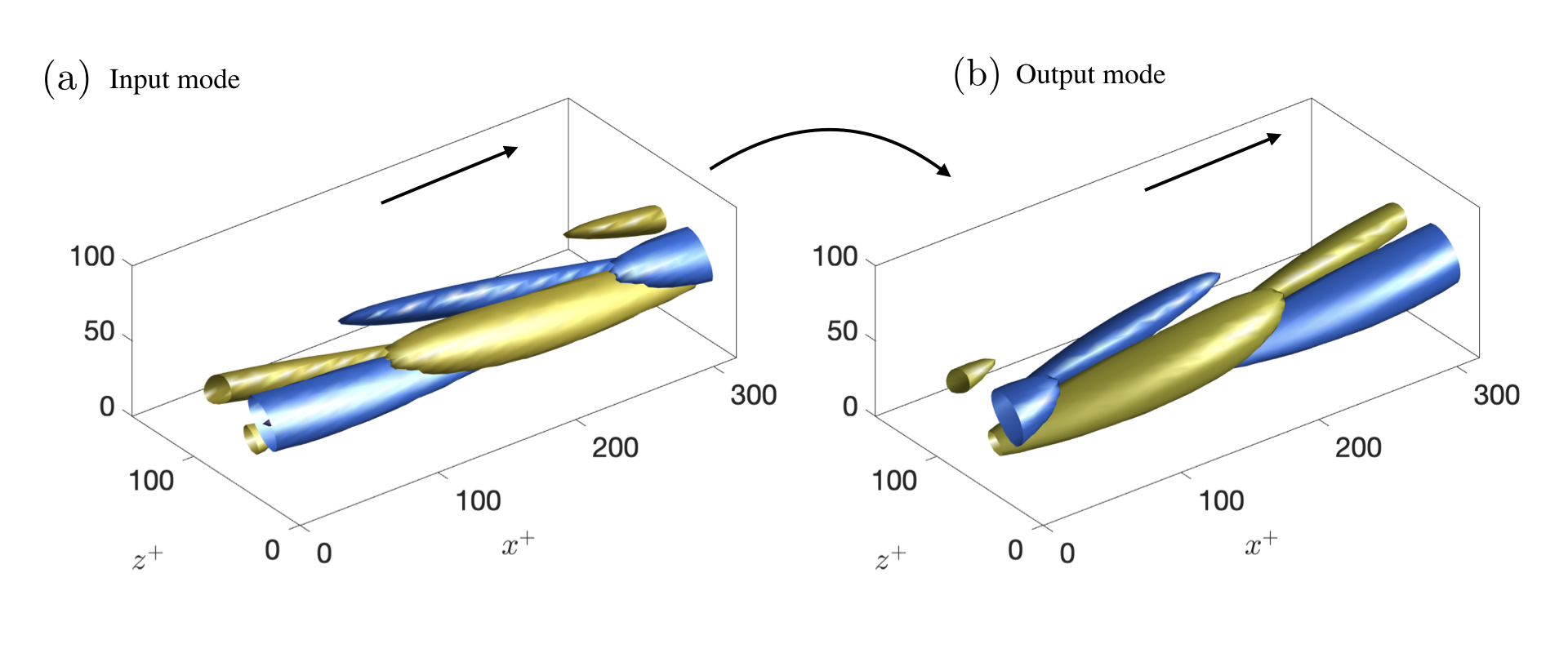}
 \end{center}
 \caption{Representative input and output modes associated with
   transient growth of the streak. Isosurfaces of (a) the input and
   (b) the output wall-normal velocity mode associated with the
   largest singular value of $\mathcal{L}(U)$ at $t=5 h/u_\tau$. The
   isosurface are $-0.5$ (blue) and $0.5$ (yellow) of the maximum
   wall-normal velocity. The gain is $G_{\mathrm{max}}=110$. The
   result is for the regular channel. \label{fig:TG_example} }
\end{figure}

Figure \ref{fig:TG_example} provides a visual representation of the
input and output modes associated with the maximum optimal gain for
one selected instant. The example displays a backwards-leaning
perturbation (input mode) tilted forward by the mean shear (output
mode).  The process is reminiscent of the linear Orr/lift-up mechanism
driven by continuity and the wall-normal transport of momentum
characteristic of the bursting process and streak formation
\cite{Orr1907, Ellingsen1975, Kim2000, Jimenez2013}.
%

\begin{figure}
 \begin{center}
   \includegraphics[width=0.75\textwidth]{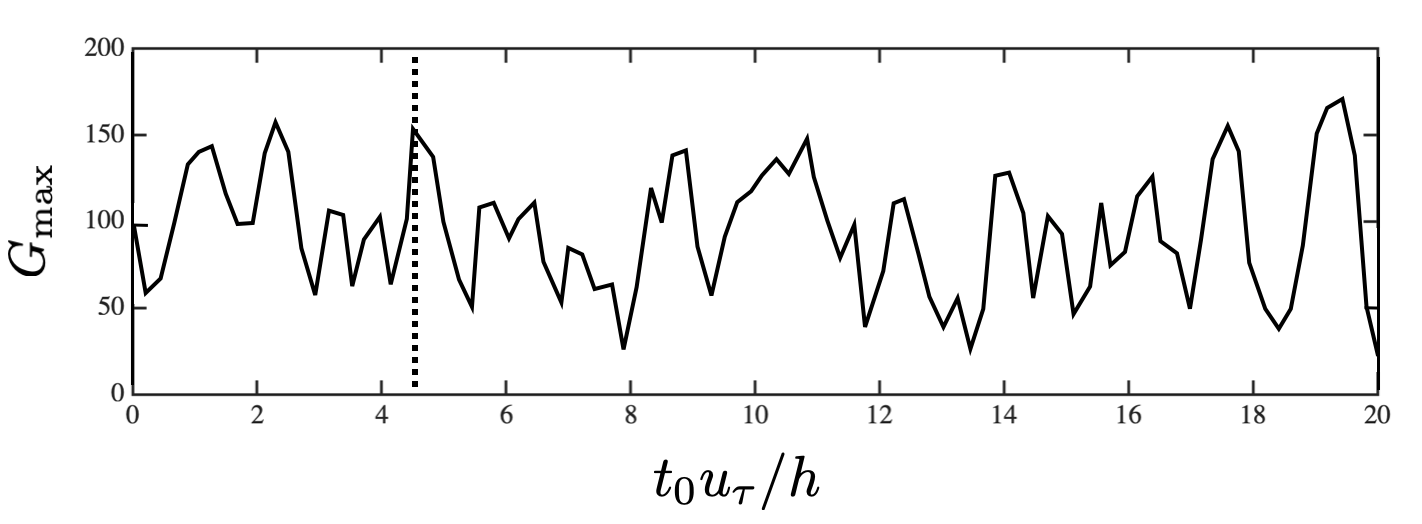}
 \end{center}
\caption{ Maximum optimal gain ($G_\mathrm{max}$) supported by
  $U(y,z,t_0)$ as a function of the fixed time $t_0$ and for a time
  horizon $T=0.5 u_\tau/h$. The results are for the channel with
  suppressed exponential instabilities. The vertical dashed line represents
  the time selected for
  Fig. \ref{fig:stats_nonmodal}.\label{fig:gains}}
\end{figure}

The maximum optimal gain for $T=0.5 h/u_\tau$ is included in Figure
\ref{fig:gains} as a function of the reference time $t_0$. The results
are for the channel with suppressed exponential instabilities, which
are on average 10\% lower than the maximum optimal gains for the
regular channel (as discussed below). The gain attains amplification
values of the input mode of the order of 100. Therefore, transient
growth supported by the ``frozen'' mean streamwise flow $U$ stands as
a tenable candidate to sustain wall turbulence. It is worth noting
that the gains $G_\mathrm{max}$ due to the non-normality of an
instantaneous mean flow with spanwise/crosstream structure $U(y,z,t)$
are remarkably large. This is in contrast with the gains traditionally
reported for the mean velocity profile defined as the average
streamwise velocity in all homogeneous directions and time, for which
$G_{\mathrm{max}}$ is limited to a more modest factor of 10
\cite{Delalamo2006a, Pujals2009}.
%

The effect of non-modal transient growth as a main source for energy
injection to the fluctuating velocities is assessed by ``freezing''
the base flow $U(y,z,t_0)$ at the instant~$t_0$. In order to steer
clear of the effect of exponential instabilities, the numerical experiment
is performed for the channel with suppressed exponential
instabilities. Simulations are then continued for $t>t_0$. This
procedure was repeated for 100 different $t_0$. The set-up disposes of
energy transfers that are due to both modal and parametric
instabilities, while maintaining the transient growth of
perturbations. The expected scenario consistent with sustained
turbulence \cite{Schoppa2002} is the non-modal amplification of
perturbations until saturation followed by nonlinear scattering and
generation of new disturbances.

\begin{figure}
 \begin{center}
   \includegraphics[width=1.0\textwidth]{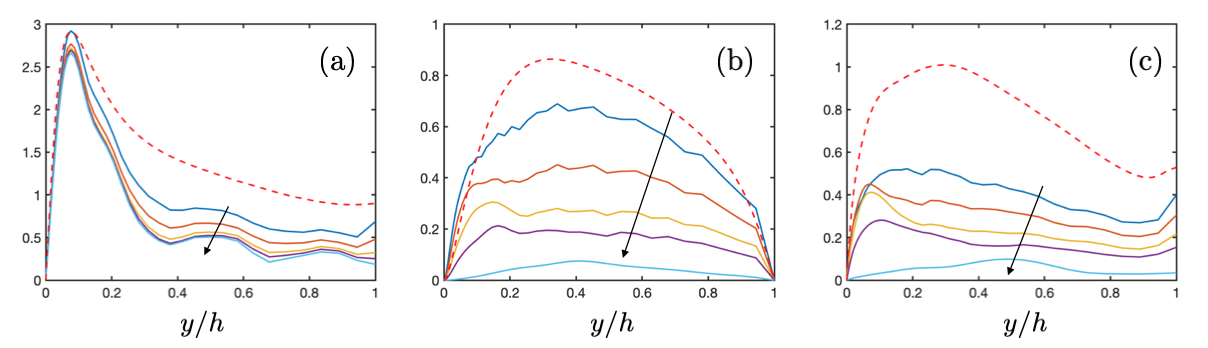}
 \end{center}
\caption{ Turbulent channel exclusively supported by transient
  growth. Root-mean-squared (a) streamwise, (b) wall-normal, and (c)
  spanwise fluctuating velocities. The dashed line is for the
  regular channel and solid lines represent different times of a
  turbulent channel exclusively supported by transient growth from
  $t=t_0$ to $t=t_0+10h/u_\tau$, where $t_0$ is the instant at which
  the mean flow is frozen in time. The arrow indicates the direction
  of time.
  \label{fig:stats_nonmodal}}
\end{figure}

The evolution of the root-mean-squared fluctuating velocities for one
of the experiments is shown in Figure~\ref{fig:stats_nonmodal}, which
contains instances from $t=t_0$ to $t=t_0+10h/u_\tau$. The particular
$t_0$ selected is highlighted in Figure \ref{fig:gains} and
corresponds to a gain of $\sim$150. After freezing the base flow,
turbulence decays and reaches a quasi-laminar state with residual
turbulence intensities required to support the prescribed
$U(y,z,t_0)$. This behavior was observed for all of the one hundred
cases investigated. Although not shown here, another hundred
additional cases were investigated again by fixing $U(y,z,t_0)$, but
initializing the flow field with random perturbations, rather than
utilizing as initial condition the already-existing fluctuating
velocities at $t=t_0$. The intensities of the random perturbations
were adjusted to be 10\% higher than their values in the regular
channel. All the additional simulations decayed similarly to the cases
reported above.  Consequently, our results suggest that turbulence is
not exclusively supported by transient growth, at least when the
magnitude of the perturbations corresponds to those typically
encountered at the low Reynolds number used here. The outcome of this
experiment does not imply that transient growth is inconsequential for
wall turbulence, but merely that additional linear ingredients are
needed to attain fully self-sustaining turbulence.
%

\begin{figure}
 \begin{center}
   \includegraphics[width=0.9\textwidth]{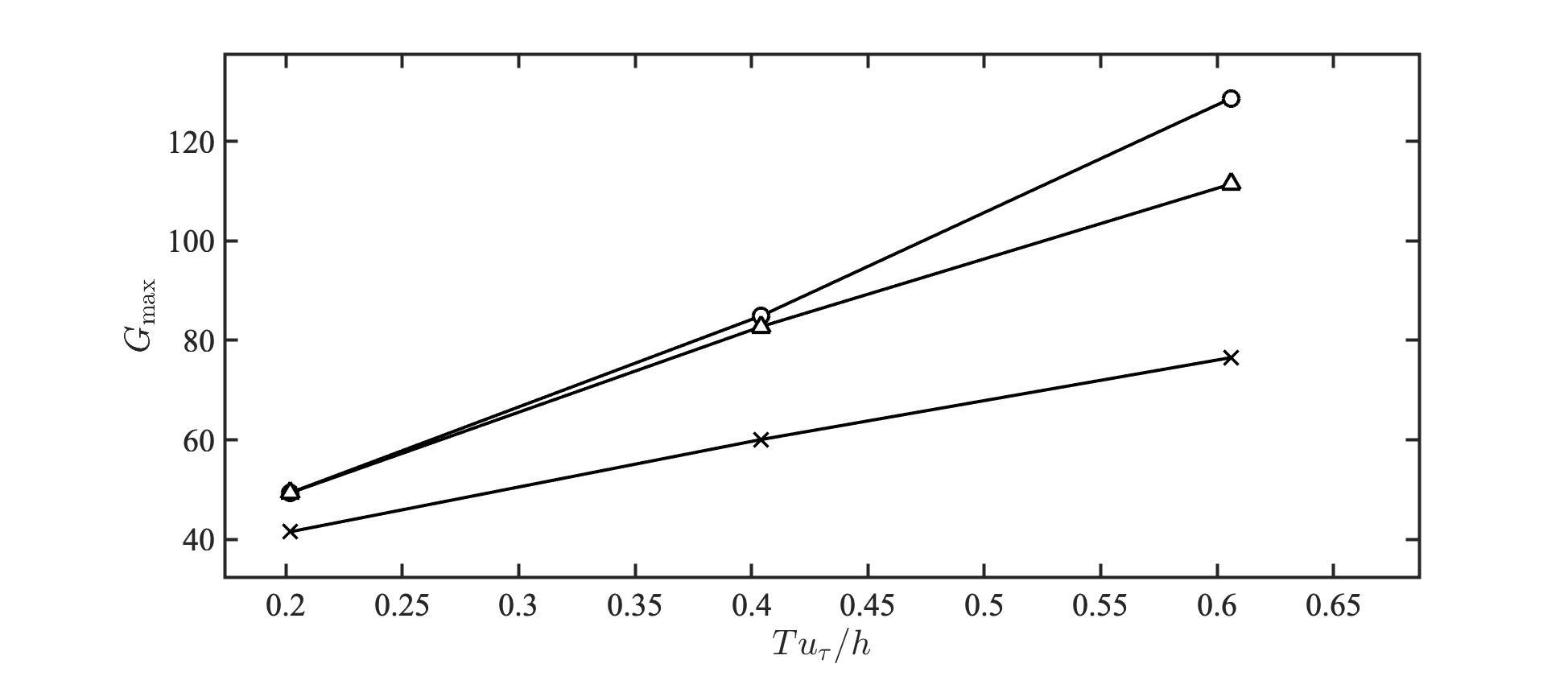}
 \end{center}
\caption{ Maximum optimal gain, $G_\mathrm{max}$, attained at the time
  $t=T$ for the regular channel (\circle), channel with suppressed
  exponential instabilities (\trian), and channel with linear damping with
  $\mu=0.43 u_\tau/h$ ($\times$). \label{fig:damped_TG}}
\end{figure}

To finalize our analysis on transient growth, we discuss briefly the
implications of stabilizing the eigenvalues of $\mathcal{L}$ as
discussed in Section \ref{subsec:nonmodal}. Due to the non-normal
nature of $\mathcal{L}$, the suppression of exponential instabilities by
Eq.~(\ref{Atilde}) inevitably entails the modification of the
non-normal characteristics of the operator. The impact on the amount
of transient growth supported by $\mathcal{L}$ is quantified in Figure
\ref{fig:damped_TG}, which shows the maximum gain of $\mathcal{L}$ and
$\tilde{\mathcal{L}}$ for different time horizons $T$. The decrease in
the maximum gain is about 15\% for $T \approx 0.6 h/u_\tau$, but
negligible for $T<0.4 h/u_\tau$. In a preliminary work
\cite{Lozano_brief_2018b}, we had proposed to incorporate the linear
friction $-\mu \boldsymbol{u}'$ into Eq.~(\ref{eq:NS}) in an attempt
to damp all the unstable eigenvalues of $\mathcal{L}$. The results in
Ref. \cite{Lozano_brief_2018b} showed that turbulence was not
sustained for the values of $\mu$ necessary to stabilize
$\mathcal{L}$. However, a detailed analysis of the impact of the
linear damping on transient growth shows that the decrease in the
gains of $\mathcal{L}$ due to the drag term $-\mu \boldsymbol{u}'$ is
proportional to $\exp(- 2 \mu T)$. Hence, values of $\mu$ necessary to
neutralize the exponential instabilities of $\mathcal{L}$ are highly
disruptive of the transient growth, which might be the cause for the
lack of sustained turbulence in Ref. \cite{Lozano_brief_2018b}. The
maximum gain of $\mathcal{L} - \mu \mathcal{I}$, where $\mathcal{I}$
is the identity operator, is also included in Figure
\ref{fig:damped_TG}.
%
 
\subsection{The parametric instability: transient growth enhancement}

The maintenance of turbulence by transient growth assisted by
parametric instability was demonstrated in Section
\ref{subsec:nonmodal}. Here, we review some aspects of the parametric
amplification mechanisms by analyzing the results from the channel
with suppressed exponential instabilities.

Parametric instability of the streak relies on an endogenous and
essentially non-modal growth process that is inherent to
time-dependent dynamical systems. The mechanism is analogous to the
classic instability of a damped harmonic oscillator with periodically
varying restoring force, in which the time-dependent variations of the
eigenbasis has the potential to altering the instantaneous linear
stability properties of the system.
%
To clarify this process, let us reformulate the linear dynamics of
Eq.~(\ref{eq:NS}) for a time period $T>0$ in terms of the propagator
$\mathcal{P}_{t \rightarrow t+T}$ as
\begin{equation}
\boldsymbol{u}'_{\mathrm{linear}}(t+T) = \mathcal{P}_{t \rightarrow t+T}
\boldsymbol{u}'(t).
\end{equation}
The propagator $\mathcal{P}_{t \rightarrow t+T}$ represents the
cumulative effect along the time $T$ of the linear operator
$\tilde{\mathcal{L}}$.  As such, it can be obtained by the discrete
approximation under the assumption of small $\Delta t$ as
\begin{equation}\label{eq:P}
\mathcal{P}_{t \rightarrow t+T} \approx \exp \left[
  \tilde{\mathcal{L}}(t+n\Delta t)\Delta t \right]\dotsb  \exp\left[
  \tilde{\mathcal{L}}(t+\Delta t) \Delta t \right] \exp\left[
  \tilde{\mathcal{L}}(t)          \Delta t \right],
\end{equation}
where $T = n\Delta t$, with $n$ a positive integer.

To evaluate the instabilities arising from the parametric mechanism,
we reconstruct the propagator for the channel with suppressed
exponential instabilities. A time-series of the base mean flow was
stored and used to estimate $\mathcal{P}_{t \rightarrow t+T}$
following Eq.~(\ref{eq:P}) with $T=h/u_\tau$. This estimation is then
used to compute the largest singular value of $\mathcal{P}_{t
  \rightarrow t+T}$, which has a value equal to $\sigma_{\mathrm{max}}
\approx 50$.  The largest singular value $\sigma_{\mathrm{max}}$ can
be compared to the non-parametric counterpart assuming a frozen mean
flow
\begin{equation}
\mathcal{P}_{t \rightarrow t+T}^0 = \exp \left[ \tilde{\mathcal{L}}(t_0)T  \right],
\end{equation}
which yields a value of $\sigma_{\mathrm{max}}^0 \approx 10$.  This
result reveals the existence of enhanced growth rates owing to the
time-variation of the streak reflected on
$\tilde{\mathcal{L}}(t)$. Moreover, for $T \rightarrow \infty$, the
growth rate of $\mathcal{P}_{t \rightarrow t+T}^0$ tends to $0$, as
the operator $\tilde{\mathcal{L}}(t_0)$ is exponentially stable. On
the contrary, a time-varying $\tilde{\mathcal{L}}$ can produce finite
exponential growth for an arbitrary $T$. It is here important to note
that if the operators $\tilde{\mathcal{L}}(t)$ at each instance were
also normal (additionally to being stable), parametric instability
would not arise. This simple but revealing example illustrates how a
time-dependent $U$ together with non-modal growth may provide the
additional energy injection into the fluctuations to attain
self-sustaining turbulence.

\section{Conclusions}\label{sec:conclusions}

We have investigated the energy injection from the streamwise-averaged
mean flow to the turbulent fluctuations. This energy transfer is
believed to be correctly represented by the linearized Navier--Stokes
equations, and three potential linear mechanisms have been considered,
namely, exponential instability of the streamwise mean $U(y,z,t)$,
non-modal transient growth, and non-modal transient growth supported
by parametric instability.

We have devised two numerical experiments of a turbulent channel flow
in which the linear operator is altered to neutralize one or various
linear mechanisms for energy extraction. In the first experiment, the
linear operator is modified to render any exponential instabilities of
the streaks stable, thus precluding the energy transfer from the mean
to the fluctuations via exponential growth. In the second experiment,
we simulated turbulent channel flows with prescribed exponentially
stable mean streamwise flow frozen in time, suppressing in that manner
any potential parametric and exponential instabilities.  Our results
establish that wall turbulence with realistic mean velocity and
turbulence intensities persists even when exponential instabilities
are suppressed. On the contrary, turbulence decays when the energy
transfer from the base flow to the fluctuating field occurs only via
transient growth.  These results argue in favor of the
parametric-instability scenario proposed by Farrell, Ioannou, and
co-workers as the basic mechanism sustaining wall turbulence
\cite{Farrell2012, Farrell2016, Farrell2017}.

Our conclusions are preliminary and refer to the dynamics of wall
turbulence in channels computed using minimal flow units, chosen as
simplified representations of naturally occurring wall turbulence.
The approach presented in this study paves the path for future
investigations at high-Reynolds-numbers turbulence obtained for larger
unconstraining domains, in addition to extensions to different flow
configurations in which the role of instabilities remains elusive.

\section*{Acknowledgements}

A.L.-D. acknowledges the support of the NASA Transformative
Aeronautics Concepts Program (Grant~No.~NNX15AU93A) and the Office of
Naval Research (Grant~No.~N00014-16-S-BA10). N.C.C.~was supported by
the Australian Research Council (Grant~No.~CE170100023). This work was
also supported by the Coturb project of the European Research Council
(ERC-2014.AdG-669505) during the 2019 Coturb Turbulence Summer
Workshop at the Universidad Polit\'ecnica de Madrid. We thank Jane
Bae, Brian Farrell, Petros Ioannou, and Javier Jim\'enez for
insightful discussions.

\section*{References}

\end{document}